\documentclass[aps,prl,twocolumn]{revtex4}
\usepackage{amsmath}
\usepackage{graphicx}
\begin{document}
\preprint{ }
\title{Theory of Optical Orientation in n-type Semiconductors}
\author{W. O. Putikka}
\affiliation{Physics Department, The Ohio State University, 1680 University Dr., Mansfield,
OH 44906}
\author{R. Joynt}
\affiliation{Physics Department, University of Wisconsin, Madison, WI 53706}

\begin{abstract}
Time resolved measurements of magnetization in n-GaAs have revealed a rich
array of spin decoherence processes, and have shown that fairly long lifetimes
($\sim100$ ns) can be achieved under certain circumstances. In time-resolved
Faraday rotation and time-resolved Kerr rotation the evolution of the
magnetization can be followed as a function of temperature, applied field,
doping level and excitation level. We present a theory for the spin relaxation
in n-GaAs based on a set of rate equations for two interacting thermalized
subsystems of spins: localized states on donor sites and itinerant states in
the conduction band. The conduction band spins relax by scattering from
defects or phonons through the D'yakonov-Perel' mechanism, while the localized
spins relax by interacting with phonons (when in an applied field) or through
the Dzyaloshinskii-Moriya interaction. In this model, numerous features of the
data, including puzzling temperature and doping dependences of the relaxation
time, find an explanation.

\end{abstract}
\volumeyear{year}
\volumenumber{number}
\issuenumber{number}
\eid{identifier}
\date{\today}
\maketitle

Spin coherence in semiconductors is attracting renewed attention due to the
prospects of spintronics -- information storage and processing using spin
rather than charge degrees of freedom, and by the idea that spins in
semiconductors could serve as qubits for quantum computers \cite{loss}%
,\cite{kane}. For qubit applications, the spin degreees of freedom must be
coherent, which necessitates a detailed understanding of the processes that
limit spin lifetimes. Time-resolved measurements on n-type systems have
revealed an array of decoherence processes, and have shown that fairly long
lifetimes (greater than $100$ ns) can be achieved in n-GaAs\cite{kikkawa98}%
,\cite{dzhioev}. In time-resolved Faraday rotation and time-resolved Kerr
rotation the evolution of the magnetization can be followed as a function of
temperature, applied field, doping level, and the intensity and duration of
the pump pulse. Results on different materials (GaAs, GaN, ZnSe) are similar,
pointing to universality in the phenomena. Awschalom and Samarth \cite{awschalom02}
have reviewed the experimental situation.

Our theory provides a systematic framework for investigating the wide range of
parameters studied in optical orientation experiments on n-type
semiconductors. Previous theory has concentrated on either higher temperatures
\cite{kikkawa98}, or on very low temperatures and very low magnetic fields
\cite{dzhioev}. We account for certain puzzling experimental observations by
having two distinct types of spin states: localized donor states and itinerant
conduction band states with characteristic spin relaxation rates $1/\tau_{l}$
and $1/\tau_{c}$, respectively. In addition, a fast cross-relaxation rate,
$1/\tau_{cr}$, between the localized and itinerant spins is a crucial feature
of these systems, leading to the largest relaxation rate, either $1/\tau_{c}$
or $1/\tau_{l}$, usually tending to dominate the spin dynamics for the whole
system. Examples of this behavior are shown in Figs. 1 and 2 below. In this
paper we focus on presenting the theory and applying it to n-GaAs, reserving
more extensive comparision to experiment for a later publication.

The cross-relaxation between localized and itinerant spins occurs by the usual
exchange interaction Hamiltonian
\[
H_{l-c}=\frac{J}{V}\sum_{i,\vec{k}}\vec{s}_{i}\cdot\vec{S}_{\vec{k}},
\]
where the sum runs over impurity spins $i$ and conduction band states $\vec
{k}$ and $V$ is the volume of the system. \ This Hamiltonian conserves total
spin and cannot, by itself, relax the magnetization. \ However, it can
transfer spin from localized to itinerant states. \ $J$ may be estimated as
$J\sim-e^{2}a_{B}^{2}.$ Here $a_{B}=10.4$ nm is the effective Bohr radius for
an impurity. If virtual excitations to the upper Hubbard band are important,
this estimate could be reduced and $J$ could even change sign. The sign is
actually not important for our purposes, since all experimental temperatures
are well above the Kondo temperature. For n-GaAs an order-of-magnitude
estimate for the cross-relaxation rate is $1/\tau_{cr}\sim(1$ ps$^{-1}%
)(n_{imp}/n_{0})$, where $n_{0}=10^{18}$ cm$^{-3}$ is a fiducial density.

The spin orientation is created by a circularly-polarized optical pump pulse
about $100$ fs long tuned near the band gap, creating particle-hole pairs.
\ The valence band holes depolarize quickly and fast recombination (on a time
scale of $50$-$100$ ps) leaves the conduction band \ and localized donor state
system with a net spin polarization along the propagation direction of the
beam ($z$-direction). \ The time evolution of this polarization is tracked by
applying a transverse magnetic field in the $x$-direction (Voigt geometry).
\ The resulting precession about the $x$-axis and concomitant decay are
measured optically, with $1/T_{2}^{\ast}$ the relaxation rate of the
macroscopic transverse magnetization.

Our theory may be formalized by writing modified Bloch equations for the
magnetization for times after recombination ($t>100$ ps after the end of the
pump pulse). \ The holes have recombined and spin-conserving processes have
thermalized the system subject to the constraint that the magnetization
retains the polarization produced by the initial excitation process. \ There
are then two thermalized subsystems of electrons at ambient temperature with
relative occupations determined by standard thermodynamic methods. The
localization of conduction band electrons onto impurity sites takes place at a
temperature scale $T_{imp}$ determined by $n_{imp}$. For $n_{imp}%
=10^{16}\text{ cm}^{-3}$ we have $T_{imp}\approx50$ K. Denote the localized
and conduction band densities by $n_{l}$ and $n_{c}$, with $n_{imp}%
=n_{l}+n_{c}$. In the experiments, $N_{ex}$, the density of electrons excited
by the pump pulse, is small, $N_{ex}\ll n_{imp}$, except possibly for
nominally insulating samples, which we discuss briefly below.

We work in the frame which rotates about the $x$-axis at a rate $g^{\ast}%
\mu_{B}B/\hbar$, and is along the $z$-axis at time $t=0$. \ In this frame, the
dynamics are governed by
\begin{align}
\frac{dn_{c+}}{dt}  &  =-\frac{n_{c+}}{\gamma_{cr}}n_{l-}+\frac{n_{c-}}%
{\gamma_{cr}}n_{l+}-\frac{1}{2\tau_{c}}n_{c+}+\frac{1}{2\tau_{c}}n_{c-}\\%
\frac{dn_{c-}}{dt}  &  =-\frac{n_{c-}}{\gamma_{cr}}n_{l+}+\frac{n_{c+}}%
{\gamma_{cr}}n_{l-}-\frac{1}{2\tau_{c}}n_{c-}+\frac{1}{2\tau_{c}}n_{c+}\\
\frac{dn_{l+}}{dt}  &  =-\frac{n_{l+}}{\gamma_{cr}}n_{c-}+\frac{n_{l-}}%
{\gamma_{cr}}n_{c+}-\frac{1}{2\tau_{l}}n_{l+}+\frac{1}{2\tau_{l}}n_{l-}\\
\frac{dn_{l-}}{dt}  &  =-\frac{n_{l-}}{\gamma_{cr}}n_{c+}+\frac{n_{l+}}%
{\gamma_{cr}}n_{c-}-\frac{1}{2\tau_{l}}n_{l-}+\frac{1}{2\tau_{l}}n_{l+},
\end{align}
where $+$ and $-$ denote up and down spins in the rotating frame and
$\gamma_{cr}=n_{0}\tau_{cr}$. 
By rewriting Eq. 1-4 in terms of the total densities, $n_l=n_{l+}+n_{l-}$ and
$n_c=n_{c+}+n_{c-}$ and the magnetization densities, $m_l=n_{l+}-n_{l-}$ and
$m_c=n_{c+}-n_{c-}$ we find that the total densities are time independent,
$dn_l/dt=0$ and $dn_c/dt=0$ and the magnetization densities are determined
by
\begin{align}
\frac{dm_c}{dt}  &  = -\left(\frac{1}{\tau_c}+\frac{n_l}{\gamma_{cr}}\right)m_c
+\frac{n_c}{\gamma_{cr}}m_l\\
\frac{dm_l}{dt}  &  = \frac{n_l}{\gamma_{cr}}m_c-\left(\frac{1}{\tau_l}+
\frac{n_c}{\gamma_{cr}}\right)m_l.
\end{align}

In general, the time dependence of the
total magnetization $m(t)=m_c(t)+m_l(t)$ is a sum of two exponentials,
$\exp{(-\Gamma_{+}t)}$ and $\exp{(-\Gamma_{-}t)}$
(behavior observed in experiments on n-GaN\cite{beschoten01}) with eigenvalues
\begin{equation}
\Gamma_{\pm}=\frac{1}{2}\left(\frac{1}{\tau_c}+\frac{1}{\tau_l}+\frac{n_{imp}}
{\gamma_{cr}}\pm S\right),
\end{equation}
where $S$ is given by
\begin{equation}
S=\sqrt{\left(\frac{1}{\tau_l}-\frac{1}{\tau_c}+\frac{n_c-n_l}{\gamma_{cr}^2}\right)^2
+\frac{4n_l n_c}{\gamma_{cr}^2}}.
\end{equation}

For n-GaAs we are in the regime $1/\tau_{cr}\gg1/\tau_{c}$, $1/\tau_{l}$, where the
eigenvalues give two very different relaxation rates: a very rapid relaxation given by
$\Gamma_{+}\approx n_{imp}/(n_0\tau_{cr})$, with a timescale on the order of picoseconds and a
slower relaxation given by
\begin{equation}
\Gamma_{-}=\frac{1}{T_{2}^{\ast}}\approx\frac{n_{l}/n_{imp}}{\tau_{l}}+\frac{n_{c}/n_{imp}%
}{\tau_{c}},
\end{equation}
with a timescale on the order of tens of nanoseconds.
Given expressions for $1/\tau_{l}$ and $1/\tau_{c}$, Eq. 9 gives the 
calculated total relaxation rate from our theory. This is the appropriate quantity
to compare to the single exponential time dependence observed in experiments\cite{kikkawa98},
\cite{dzhioev}. 
\begin{figure}[ptb]
\includegraphics[width=3.4in,clip]{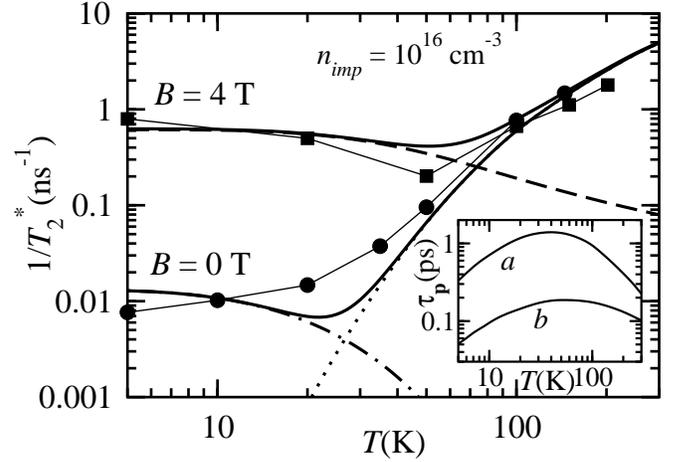}
\caption{Plot of $1/T_{2}^{\ast}$ versus temperature. The data are from Ref.
\onlinecite{kikkawa98}, with solid dots for $B=0$ T and solid squares for
$B=4$ T. The lines connecting the data points are guides for the eye. The
curves are a least-squares fit of Eq. 5 to the data. Dashed-dotted curve:
$(n_{l}/n_{imp})(1/\tau_{DM})$, Dashed curve: $(n_{l}/n_{imp})(1/\tau_{s-ph})$
for $B=4$ T, Dotted curve: $(n_{c}/n_{imp})(1/\tau_{c})$, Solid curves: total
$1/T_{2}^{\ast}$. For $B=0$ T, $1/\tau_{s-ph}=0$. Inset: $\tau_{\mathbf{p}}$
versus temperature for $a$: $n_{imp}=10^{16}\text{ cm}^{-3}$ and $b$:
$n_{imp}=10^{18}\text{ cm}^{-3}$.}%
\end{figure}

There are various processes that can relax the non-equilibrium magnetization
produced in optical orientation experiments. The conduction band processes
have been well studied\cite{pikus}, while the relaxation mechanisms for
localized electrons are less well understood.

\textbf{D'yakonov-Perel' Mechanism} Conduction band electron spins in n-GaAs
relax primarily by the D'yakonov-Perel' (DP) mechanism \cite{Dyakonov74}, due
to lack of inversion symmetry in III-V systems. Lack of inversion symmetry,
together with spin-orbit coupling, gives an effective $\vec{k}$-dependent
magnetic field, causing the spin of an itinerant electron to precess about an
axis related to $\vec{k}$. \ The precession frequency for an electron at
wavenumber $\vec{k}$ is $\Omega_{DP}(\vec{k})$, and the DP relaxation comes
from switching the precession axis by scattering from one $\vec{k}$-vector to
another. \ Assuming s-wave scattering, one obtains $1/\tau_{DP}(\vec
{k})=2\Omega_{DP}^{2}(\vec{k})\tau_{p}(\vec{k})/3$, where $\tau_{p}(\vec{k})$
is the momentum relaxation time. \ Averaging this expression over the
Boltzmann distribution, and using the results of Fishman and Lampel for the
momentum average \cite{fishman}, we find a spin relaxation rate $1/\tau
_{DP}=\alpha_{DP}T^{3}\tau_{p},$ with $\alpha_{DP}($th$)=9.0\times10^{-10}$
K$^{-3}$ps$^{-2}$. Here $\tau_{p}$ is the average momentum relaxation time
which has a complicated temperature and doping dependence (shown in Fig. 1)
best taken from mobility data, $\mu_{e}=e\tau_{p}/m^{\ast}$ \cite{ulbrich00}.

\textbf{Elliot-Yafet Mechanism} Conduction band electron spins can also relax
via the Elliot-Yafet (EY) \cite{Elliott54} mechanism, due to ordinary impurity
scattering from state $\vec{k}$ to state $\vec{k}^{\prime}$. With spin-orbit
coupling the initial and final eigenstates are not eigenstates of $S_{z}$, the
spin projection operator, so this process relaxes the spin. \ One finds
$1/\tau_{EY}=\alpha_{EY}T^{2}/\tau_{p}$, where 
$\alpha_{EY}(\text{th})=8\times10^{-10}$ K$^{-2}$.
For the experimental parameters discussed below $(n_{c}/n_{imp})(1/\tau_{EY})$
is three orders of magnitude smaller than the leading contributions to 
$1/T_{2}^{\ast}$. Thus, to simplify
our analysis of experiments on n-GaAs, we set $\alpha_{EY}=0$ and $1/\tau
_{c}=1/\tau_{DP}$.

\begin{figure}[ptb]
\includegraphics[width=3.4in,clip]{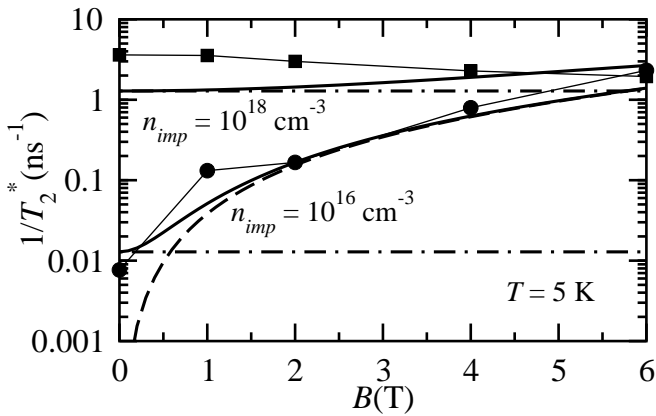}
\caption{Plot of $1/T_{2}^{\ast}$ versus applied magnetic field. The data are
from Ref. \onlinecite{kikkawa98}, with solid dots for $n_{imp}=10^{16}%
\text{\ cm}^{-3}$ and solid squares for $n_{imp}=10^{18}\text{\ cm}^{-3}$. The
lines connecting the data points are guides for the eye. The curves are a
least-squares fit of Eq. 5 to the data. Dashed-dotted curves: $(n_{l}%
/n_{imp})(1/\tau_{DM})$ for the two densities, Dashed curve: $(n_{l}%
/n_{imp}(1/\tau_{s-ph})$, Solid curves: total $1/T_{2}^{\ast}$. For $T=5$ K,
$(n_{c}/n_{imp})(1/\tau_{c})\ll0.001$ ns$^{-1}$.}%
\end{figure}
Spins localized on
donor sites cannot relax by the same scattering dependent processes that relax
conduction band spins. As a rule, relaxation times for localized states are
longer than for itinerant states, due to phase space effects. Lifetimes for
localized states can be very long: times in excess of $10^{3}$ s have been
measured for donor bound states of phosphorus-doped silicon \cite{feher}. \ 

\textbf{Spin-phonon Mechanism}
Acoustic phonons can relax localized spins by dephasing, due to
spin-orbit coupling mixing spin states, if an external field $B$ breaks the
time-reversal symmetry present at zero field (Van Vleck cancellation). With $B$ in the
$z-$direction, the relevant term in the Hamiltonian for a single spin takes
the form%
\begin{equation}
H_{s-ph}=C\mu_{B}B\sigma_{zi}\Delta_{i},
\end{equation}
where $\Delta_{i}$ is the dilatation at a donor site $i$ and $C$ is a constant. \ The
dilatation modulates $g^{\ast},$ the effective $g$-factor, which is given by
$g^{\ast}=2[1-(m^{\ast}/m-1)\Delta_{so}/(3E_{g}+2\Delta_{so})],$ where
$E_{g}=1.4$ eV is the energy gap, $m^{\ast}/m=0.067$ is the ratio of the
effective mass to the bare mass, and $\Delta_{so}=0.344$ eV is the spin-orbit
splitting of the valence bands \cite{roth}. \ We also have that $E_{g}%
(\Delta_i)=E_{g}(0)-(9\text{eV})\Delta_i$ and the effective mass depends on the
gap as $m^{\ast}\sim1/E_{g}.$\ These facts allow us to estimate $C^{2}%
\approx550$. \ There is no generally accepted theory of the multi-spin
relaxation rate $1/T_{2}^{\ast}$ that results from such a Hamiltonian (the
single-spin rate $1/T_{2}$ has recently been calculated in silicon
\cite{mozyrsky}). We can obtain a simple estimate using Redfield theory, which
yields $1/\tau_{s-ph}=C^{2}B^{2}\left\langle \Delta^{2}\right\rangle \tau
_{ph},$ where $\tau_{ph}$ is the phonon correlation time \cite{slichter}
and $\Delta$ is the average dilatation for the occupied donors.
\ This leads finally to $1/\tau_{s-ph}=\alpha_{s-ph}B^{2}T^{4}f(T),$ where
$f(T)=\int_{0}^{\theta_{D}/T}x^{3}\left[  \left(  1/2\right)  +(e^{x}%
-1)^{-1}\right]  dx$ in a Debye model for the phonons and $\theta_{D}=343$ K
for GaAs. We note that this theory is not likely to be valid at
higher temperatures, where multi-phonon and Orbach processes become important.
\ This issue is not settled even in insulators, and we defer
full consideration of it to a later publication.

\textbf{Nuclear Hyperfine Mechanism} A localized electron spin can relax
through the hyperfine interaction with the $N\approx10^{5}$ nuclei
with which it is in contact. \ The nuclei are randomly oriented under most
conditions, and the associated field felt by the electron is
$A/\sqrt{N}g\mu_{B}\approx10^{-2}$ T, where $A$ is the hyperfine constant.
\ The corresponding precession frequency is $\omega_{N}\approx10^{8}$ s$^{-1}%
$. \ There is also a characteristic time for the nuclei $T_{n2}\approx10^{-4}$
s, which comes from the nuclear dipole-dipole interaction. \ Hence $\omega
_{N}T_{n2}\approx10^{4}$, and we are in the regime where the effective random
field fluctuates slowly compared to the precession of the spin. The relaxation
time from coupling to the nuclei, $1/\tau_{nuc}$, is not expected to have
strong temperature dependence in the range $T>1$ K considered here and there
should be no field dependence as long as $B>10^{-2}$ T. We
treat $1/\tau_{nuc}$ as a constant.

\textbf{Dzyaloshinskii-Moriya Mechanism} Localized electron spins can relax by
the Dzyaloshinskii-Moriya (DM) interaction\cite{kavokin}. This interaction,
arising from spin-orbit coupling, produces a term proportional to $\vec
{b}\cdot\vec{s}_{1}\times\vec{s}_{2}\,\ $where $\vec{b}$ is related to the
interspin separation, and to the exchange integral between the wavefunction on
sites 1 and 2. \ This interaction is not isotropic in spin space and can
therefore relax the spins. \ The calculation of the effect of this term on
$1/T_{2}^{\ast}$ is not straightforward, since it involves aspects of the spin
glass problem that are not entirely solved. \ Gor'kov and Krotkov
\cite{gorkov} have given the first term in a density expansion. \ We use their
expression, though with a more general distance dependence for the exchange
interaction \cite{landau} as a first step toward a theory valid at higher
impurity densities. We find $1/\tau_{DM}=\alpha_{DM} n_{imp}a_{B}^{3}%
f_{DM}(T)$. Here $\alpha_{DM}(\text{th})=0.01$ ns$^{-1}$ and the weakly
temperature dependent, dimensionless function $f_{DM}(T)\approx32$ at $T=5$ K.

The total relaxation rate for the localized spins in our theory is given by $1/\tau
_{l}=1/\tau_{s-ph}+1/\tau_{nuc}+1/\tau_{DM}$.

In Figs. 1 and 2 we compare the results of our theory to experimental data on
n-GaAs at $n_{imp}=10^{16}$ cm$^{-3}$ and $n_{imp}=10^{18}$ cm$^{-3}$. \ Our
procedure is as follows. \ Each mechanism above has very definite field,
temperature, and doping dependence. \ The overall constant factor for each is
less certain. The data are consistent with $1/\tau_{nuc}=0$. However, for
reasons explained below, the data do not set tight limits on $1/\tau_{nuc}$.
We do a least-squares fit to the complete data set shown in Figs. 1 and 2
using the three remaining adjustable parameters, with the results $\alpha
_{DP}(\text{exp})=8.6\times10^{-10}$ K$^{-3}$ps$^{-2}$, $\alpha_{s-ph}%
(\text{exp})=2.2\times10^{-11}$ T$^{-2}$K$^{-4}$ns$^{-1}$, and $\alpha
_{DM}(\text{exp})=0.031$ ns$^{-1}$. These values are in satisfactory agreement
with $\alpha_{DP}(\text{th})=9.0\times10^{-10}$ K$^{-3}$ps$^{-2}$ and
$\alpha_{DM}(\text{th})=0.01$ ns$^{-1}$. The remaining value $\alpha_{s-ph}%
($exp$)$ is best viewed as an estimate of the phonon correlation time,
$\tau_{ph}=1.2\times10^{-5}$ ns. This seems reasonable, given that the inverse
Debye frequency $\hbar/k_{B}\theta_{D}=2.2\times10^{-5}$ ns. But true
comparison of theory and experiment for this prefactor awaits a more
comprehensive theory of the phonon relaxation, as noted above.

The fits against temperature at $n_{imp}=10^{16}$ cm$^{-3}$ for $B=0$ T and
$B=4$ T are shown in Fig. 1. There are two surprising points about the data:
(1) $1/T_{2}^{\ast}$ is independent of $B$ at high $T$; (2) the $T$ dependence
is non-monotonic at higher fields. Point (1) is explained by noting that in
our theory all $B$ dependence comes from $1/\tau_{s-ph}$. For this doping, the
localized states are completely depopulated at high $T$. \ Point (2) is more
subtle. In our theory, though both $1/\tau_{c}$ and $1/\tau_{l}$ increase with
increasing $T$, their contributions to $1/T_{2}^{\ast}$ are weighted by
$n_{c}/n_{imp}$ and $n_{l}/n_{imp}$, respectively. Starting at low $T$, where
localized states dominate, we have $1/T_{2}^{\ast}$ decreasing with increasing
$T$ due to the decrease of $n_{l}/n_{imp}$ as $T$ increases. After the
localized states are depopulated $1/\tau_{c}$ dominates and $1/T_{2}^{\ast}$
is an increasing function of $T$.

The fits against field for $n_{imp}=10^{16}$ cm$^{-3}$ and $n_{imp}=10^{18}$
cm$^{-3}$ at $T=5$ K are shown in Fig. 2. \ These results are also surprising:
there is strong enhancement of $1/T_{2}^{\ast}$ by $B$ at low doping, while at
high doping, the dependence is quite weak. \ In this case, the explanation
relies on the DM contribution. \ At high doping $1/\tau_{DM}$ dominates
because of the short impurity-impurity spacing and consequent fast relaxation.
\ This rate is field-independent. \ At low doping $1/\tau_{s-ph}$ with its
strong $B$ dependence is more important.

Experiments\cite{kikkawa98,dzhioev} show $n_{imp}=10^{16}$ cm$^{-3}$ is the
\textquotedblleft optimal\textquotedblright\ (smallest $1/T_{2}^{\ast}$)
doping value at low $T$. \ Going to lower dopings (data not shown) increases
$1/T_{2}^{\ast}$. \ This is due to a combination of effects. \ Increasing the
doping at first raises the number of localized states, which have
intrinsically much smaller decay rates. \ However, this process stops when the
impurity states begin to overlap, and $1/\tau_{DM}$ dominates. \ This is why
the minimum $1/T_{2}^{\ast}$ is near the metal-insulator transition. \ These
conclusions are consistent with those of Ref. \cite{dzhioev} at small $B$ and
$T$. \ 

From an examination of Fig. 2, we can see why adding $1/\tau_{nuc}$ does not
significantly improve or worsen the fit. Doing so adds a constant to
$1/T_{2}^{\ast}$, which would move all theory curves rigidly upward, not much
affecting the overall goodness of fit. \ Thus we cannot get a meaningful limit
on $1/\tau_{nuc}$ with this data set. The question of how to average over
nuclear degrees of freedom to find $1/\tau_{nuc}$ is not completely clear, and
there are different results in the recent literature \cite{khaetskii02},
\cite{de sousa02}.

The main shortcoming in our theory lies in the high doping regime, where
spin-glass effects become important. \ This is shown by the $B=0$ points in
Fig. 2, where $1/T_{2}^{\ast}$ is dominated by $1/\tau_{DM}$. \ Since we use a
low-density (pairwise correlations only) expansion for $1/\tau_{DM}$, the
ratio of the theory points is exactly the ratio between the densities, which
is too small compared with experiment. A better theory would approach freezing
as a collective, not pairwise, effect. \ Apart from this, all the qualitative
features of the data are explained in our picture. \ \ 

Our analysis shows that the complicated $B$ and $T$ dependences for
$1/T_{2}^{\ast}$ observed in experiments are due to having two strongly
interacting subsystems of spins: one localized and the other itinerant. The
coupling of localized spins to phonons gives rise to the unusual magnetic
field depndences of the relaxation rates. \ Overall, the very different $T$
and $B$ dependences for $1/\tau_{c}$ and $1/\tau_{l}$ coupled with population
effects give the wide range of experimental phenomena observed. \ 

%\begin{acknowlegments}
This work was supported by the NSF Materials Theory program, Grants
DMR-0105659 (WOP) and DMR-0081039 (RJ). \ We are grateful to D. L. Huber and
J.W. Wilkins for helpful comments on the manuscript. \
%\end{acknowlegments}

\end{document}